\documentclass[preprintnumbers,aps,twocolumn,floatfix,superscriptaddress]{revtex4-1}
\usepackage{eso-pic,calc}
\usepackage{graphicx}
\usepackage{amsmath, amsthm, amssymb}
\usepackage{epsfig}
\usepackage{bm}
\usepackage{color}

\usepackage[colorlinks=True]{hyperref}  
\hypersetup{
	colorlinks=true,  
	linkcolor=blue,  
	citecolor=blue, 
	filecolor=magenta,   
	urlcolor=blue         
}

\def\bc{\begin{center}}
\def\ec{\end{center}}

\begin{document}
\title{Continuous sample space reducing stochastic process}

\author{Rahul Chhimpa}
\affiliation{Department of Physics, Institute of Science,  Banaras Hindu University, Varanasi 221 005, India}

\author{Avinash Chand Yadav\footnote{jnu.avinash@gmail.com}}
\affiliation{Department of Physics, Institute of Science,  Banaras Hindu University, Varanasi 221 005, India}

\begin{abstract}
We propose a simple model for sample space reducing (SSR) stochastic process, where the dynamical variable denoting the size of the state space is continuous. In general, one can view the model as a multiplicative stochastic process, with a constraint that the size of the state space cannot be smaller than a visibility parameter $\epsilon$. We study the survival time statistics that reveal a subtle difference from the discrete version of the process. A straightforward generalization can explain the noisy SSR process, characterized by a tunable parameter $\lambda \in [0, 1]$. We also examine the statistics of the size of the state space that follows a power-law distributed probability $\mathbb{P}_{\epsilon}(z\le \epsilon) \sim z^{-\alpha}$, with a nontrivial value of the exponent as a function of the tunable parameter $\alpha = 1+\lambda$. 
\end{abstract}

\maketitle

\section{Introduction}
The evolving sample space emerges in diverse dynamical processes~\cite{Murtra_2015, S_Thurner_2018}. Typical examples range from sentence formation~\cite{Murtra_2015} to innovation~\cite{novel_1} and biological extinction to fragmentation~\cite{Krap_1994, Dhar_2015, Domokos_2015, Krapivsky_2011}. Such processes are path-dependent, typically modeled as driven-dissipative systems. In this context, the {\it sample space reducing} (SSR) process~\cite{Murtra_2015} has been introduced as an alternative mechanism for the scale-invariant features observed in complex systems~\cite{Newman_2005, Clauset_2009, 2012_Porter, Reed_2002, gros_2014, Munoz_2018, Yadav_2020, Sabin_2022}. The other notable mechanisms for the origin of power-law are self-organized criticality~\cite{Bak_1987, Bak_1996, Dhar_2006, Yadav_2012}, preferential attachment~\cite{Albert_1999}, and constrained multiplicative stochastic process~\cite{Sornette_1997, Takayasu_1997, Susanna_1999, Yamamoto_2012, Yamamoto_2014}.

For the SSR, the accessible sample space stochastically decreases with time. In other words, the process gets constrained when the dynamics unfold. It is a signature of the aging effect. On the other hand, a generalized case {\it noisy} SSR with a parameter $\lambda$ shows occasional expansion. An important aspect of the process is its analytical tractability. One of the key observables is the visitation frequency $P(i)$: How frequently a state labelled as $i \in [1, N]$ is visited by the process. This follows the Zipf's law~\cite{Zipf_1949, Moreno_2016, marzo_2021} describing frequency versus rank behavior: $P(i)\sim i^{-\lambda}$ with $\lambda\in [0, 1]$. The case $\lambda = 1$ corresponds to the original SSR.

An alternative way to express Zipf's law is to examine the probability distribution of the frequency, which behaves as a decaying power-law with an exponent $\alpha = 1 +1/\lambda$. The exponent has a wide range $\alpha \in [2, \infty)$ for the parameter $\lambda$. However, for a physical observable satisfying a power-law distribution, the normalization condition puts a constraint that $\alpha>1$. Physically, the exponent value in the range 1 to 2 is significantly relevant. The SSR, after including a cascading feature, can explain the full spectrum of scaling exponents~\cite{Murtra_2017}. 
It is also interesting to mention that the variants of the SSR can explain not only the power-law distribution, but also a variety of stationary distributions by controlling the level of state-dependent noise~\cite{Thurner_2018}.

Complex systems have a modular structure. For example, texts are composed of words. The SSR appears to be a more effective generative process that can explain other statistical properties, such as the statistics of shared components and Heap's law (the number of different components as a function of system size)~\cite{Mazzolini_2018, andrea2018}.

In addition to the visitation frequency, another relevant observable for the SSR is the life span or survival time. The survival time statistics for the SSR~\cite{Yadav_2016_ssr1} have revealed a map with the record statistics for independent and identically distributed (iid) random variables. A conjecture also suggests a map between the survival time statistics for the noisy SSR and the records statistics for correlated random events~\cite{Yadav_2017_ssr2}. The relevant examples of correlated events include a drifted random walk with Cauchy distributed steps and fractional Brownian motion.

In this paper, we present a continuous version of the SSR model inspired by the description of fragmentation events. Previous works have examined a continuous SSR, focusing on the visitation frequency~\cite{Murtra_2016}. However, the treatment as well as the focus of our model differs significantly. Here, we study the survival time statistics, which remain well understood for the discrete version of the process. More interestingly, the model also represents a constrained multiplicative stochastic process. Consequently, the size of the sample space exhibits a scaling behavior, with a nontrivial value of the exponent. We apply scaling methods to examine the statistics of the sample space size. In the regime $z\le \epsilon$, the size of state space follows a power-law distribution $\mathbb{P}_{\epsilon}(z) \sim z^{-(1+\lambda)}$, where the exponent takes a value in the range 1 to 2. We verified our results by simulation studies.

The paper is structured as follows: Section~\ref {sec_2} introduces the model definition with a continuous state space. Section~\ref{sec_3} presents results for the survival time statistics in the continuous SSR, and Section~\ref {sec_4} mentions the results for the noisy SSR. The probability density for the sample space size is computed in Sec.~\ref{sec_5}. Finally, the paper is summarized in Sec.~\ref{sec_6}. An appendix shows additional analytical results.

\section{Model Definition}{\label{sec_2}}
We begin by first recalling the definition of the discrete version of the process. For an SSR stochastic process, the size of the state space decreases monotonically with time. Let us express the time evolution of the size of the sample space, $z(t)$, as a map
\begin{equation}
z(t+1) = G[z(t)],
\label{eq_g_ssr}
\end{equation}
where the function $G$ in Eq.~(\ref{eq_g_ssr}) describes the dynamical update rules. In the discrete version, both $z$ and $t$ are discrete. The update rules are the following \cite{Yadav_2016_ssr1}: For all time $0\leq t \leq \tau$, $z(t+1)<z(t)$.  $z(t+1)$ is selected randomly with uniform distribution in the interval $[1, z(t)-1]$. The process also represents a directed random hopping on $\mathbb{Z}^+$, the set of positive integers, with boundary conditions $z(0) = N$ and $z(\tau) = 1$. The process stops when $z(\tau) = 1$, where $1\leq\tau \leq N$. Here $\tau$ is a random variable denoting the survival time of the process. When the system has one state, there is no change in the state of the system; that is, there is no dynamics.

\begin{figure}[]
	\centering
	\scalebox{0.6}{\includegraphics{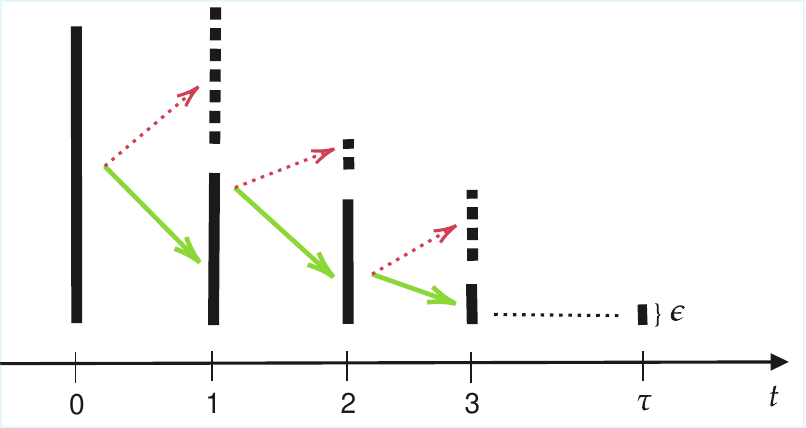}}
	\caption{The diagram depicts the dynamics of continuous SSR as a fragmentation process. A stick of unit length is considered at time $t=0$ and divided into two pieces. The solid (green) arrow shows the part selected for further update, and the dashed (red) arrow demonstrates the excluded part. The process repeats until the survival time $t=\tau$, where the length of the selected part is larger than or equal to the visibility parameter $\epsilon$.} 
	\label{fig0}
\end{figure}

The continuous version of the SSR process (cf. Fig.~\ref{fig0}) is a mathematical model, where the dynamical variable is a real number, but time may be discrete or continuous. The model's motivation comes from the description of a physical event modeled as a product of several independent random events with an additional constraint. The constraint imposed here is such that the process stops when it reaches a threshold. We iterate the same dynamics. The renewal property introduces stationarity and is crucial for the emergence of a power-law distributed size distribution (cf. Sec.~\ref{sec_4}), rather than a log-normal distribution.

When an explosion occurs, a large object is fragmented into smaller pieces, successively, in a very short time. With this observation, we can follow the SSR process in terms of fragmentation. Consider a stick of unit length. Divide it into two segments randomly with uniform distribution. Take one segment and repeat the dynamics (dividing) similarly. This process will continue as long as the object is large enough to be visibly seen. Or when the size becomes smaller than a resolution parameter, the fragmentation stops. A straightforward generalization of the discussion can help us to understand the SSR from the fragmentation of extended objects.

Now, we introduce the definition of the continuous SSR. Consider iid random variables with a probability density $p(x)$. For simplicity, let $p(x)$ be uniform, with $x \in [0,1]$. Define an event
\begin{equation}
z(t) = \prod_{i=0}^{t}x(i),
\label{zt_update}
\end{equation}
with an initial condition $z(0) = 1$.  Within the SSR framework, $z$ represents the region of accessible state space. Here, time $t$ is a discrete variable running from 0 to $\tau$ such that the state variable is above a threshold $z(t) > \epsilon$, for $0\leq t <\tau$. The process stops when $z(\tau)\leq\epsilon>0$. We can view the threshold $\epsilon$ as a visibility or resolution parameter. The process executes as long as the variable $z$ is visible, or from the SSR process point of view, it does not reach the absorbing state. The dynamics settle down when trapped in an attractor, and the parameter $\epsilon$ represents the width of the attractor.

$z(t)$ is a random variable that reduces as a function of time, monotonically. However, the visibility parameter limits the reduction in $z$. If $\epsilon = 1/N$, then we can see a clear analogy with the discrete version of the SSR. Thus, the survival time $\tau \in [1, \lfloor1/\epsilon\rfloor]$ is one of the quantities of interest. Moreover, the smaller the value of $\epsilon$ is, the longer the mean survival time.

\begin{figure}[t]
  \centering
  \scalebox{1}{\includegraphics{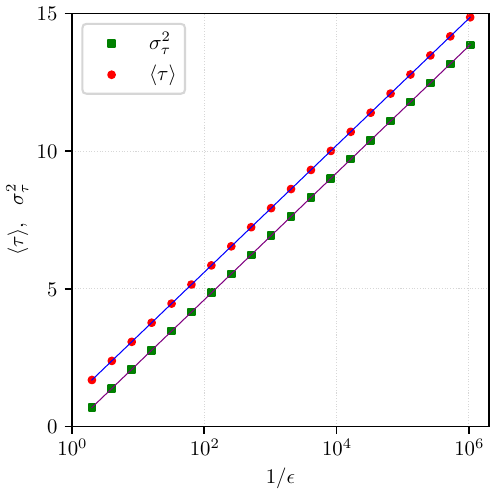}}
  \caption{For the continuous SSR: The plots show the mean and variance of the survival time as a function of the visibility parameter $\epsilon$. Each point is obtained by averaging over $10^6$ independent realizations of the SSR process. We choose the parameter $1/\epsilon$ in steps $2^i$, where $i$ runs from 2 to 20. Here, $\Delta = \langle \tau \rangle - \sigma_{\tau}^{2} \approx 1.005$ for large $N = 2^{20}$. The continuous straight lines correspond to the theoretical function [cf. Eq.~(\ref{msd1})].}
\label{fig1}
\end{figure}

\section{Survival time statistics for the continuous SSR}{\label{sec_3}}
The survival time statistics for the discrete version of the SSR remain well understood and can serve as a benchmark for the validity of a new model. Exact analytical results provided in~\cite{Yadav_2016_ssr1} suggest that  the first two cumulants for the survival time behave as
\begin{eqnarray}
\langle \tau \rangle = 1-\ln\epsilon~~~~~{\rm and}~~~~
\sigma_{\tau}^2 = -\ln \epsilon,
\label{msd1}
\end{eqnarray}
and the survival time distribution has a form
\begin{equation}
\mathcal{P}_{\epsilon}(\tau) = \frac{\mu^{\tau-1}\exp(-\mu)}{(\tau-1)!}.
\label{eq_sur_dist_ssr}
\end{equation}
For small value of $\epsilon$, the distribution is Poisson with parameter $\mu = \langle \tau \rangle \approx \sigma_{\tau}^{2} = -\ln \epsilon$. However, the distribution shown in Eq.~(\ref{eq_sur_dist_ssr}) is log-Poisson in the parameter $\epsilon$. Both mean and variance vary logarithmically as a function of $1/\epsilon$ [cf. Eq.~(\ref{msd1})].

Moreover, a map between the survival time statistics for the SSR and the records statistics for iid random variables suggests that  the survival time distribution, in the continuum limit, is Gaussian  
\begin{equation}
\mathcal{P}_{\epsilon}(\tau) \approx \frac{1}{\sqrt{2\pi \sigma_{\tau}^2}}\exp\left[-\frac{(\tau-\langle \tau \rangle)^2}{2\sigma_{\tau}^2}\right].
\label{gauss}
\end{equation}
The mean and variance behave, more precisely,
\begin{eqnarray}
\langle \tau \rangle = -\ln \epsilon  + \gamma + \mathcal{O}(\epsilon),\nonumber\\ 
\sigma_{\tau}^{2} = -\ln \epsilon + \gamma -\zeta(2) + \mathcal{O}(\epsilon),
\label{msd}
\end{eqnarray}
where $\zeta(2) = \pi^2/6$. Here $\Delta = \langle \tau \rangle - \sigma_{\tau}^{2} = \zeta(2) \approx 1.646$.

\begin{figure}[t]
  \centering
  \scalebox{1}{\includegraphics{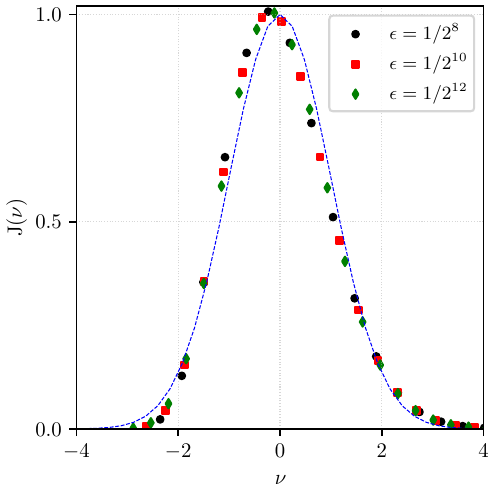}}
  \caption{The data collapse curves for the survival probability distribution, with different $\epsilon$. Here $v = (\tau - \langle \tau \rangle )/\sigma_{\tau}$ and $J(v) = \sqrt{2\pi\sigma_{\tau}^{2}}\mathcal{P}_{\epsilon}(\tau)$. To see the deviation between simulation results and the asymptotic distribution, Eq.~(\ref{gauss}) is plotted for $1/\epsilon = 10^8$ (see broken line).}
\label{fig2}
\end{figure}

We highlight an interesting remark on the difference between the mean and variance of the survival time. For the continuous SSR, the numerical results shown in Fig.~\ref{fig1} suggest $\Delta = 1$, consistent with the exact results [cf. Eq.~(\ref{msd1})]. To obtain the exact results~\cite{Yadav_2016_ssr1}, the assumptions made are that the size of the state space is continuous, and $\tau$ is discrete. But the asymptotic results show $\Delta = \zeta(2)$  [cf. Eq.~(\ref{msd})], which remains observed for the discrete version of the SSR~\cite{Yadav_2016_ssr1}. The asymptotic results are valid when we treat even $\tau$ as a continuous variable. Now it becomes clear the extent of exact and asymptotic results for the survival time statistics for the SSR. As shown in Fig.~\ref{fig2}, the shifted and scaled survival time follows a Gaussian distribution with a slight deviation, as expected.

\begin{figure}[t]
  \centering
  \scalebox{1}{\includegraphics{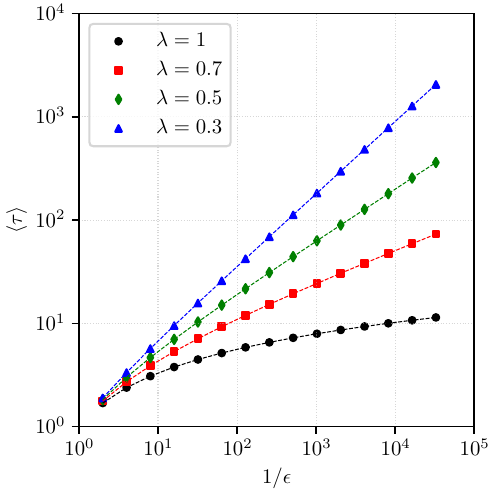}}
  \caption{For the continuous noisy SSR: The average survival time as a function of the visibility parameter for different values of $\lambda$. The maximum run time for each independent realization is $10^4$, and the curve is averaged over $10^6$ different realizations. For $\lambda = 0.3, 0.5,$ and 0.7, the corresponding estimated exponents are $\theta = 0.697(1), 0.505 (2)$, and $0.326(5)$, respectively.}
\label{fig3}
\end{figure}

\section{Survival time statistics for the noisy SSR}{\label{sec_4}}
Consider an unconstrained stochastic process
\begin{equation}
z_{t+1} = G_0(z_t) =  x(t).
\label{eq_unconst}
\end{equation} 
Then, the continuous noisy SSR is easy to define using Eq.~(\ref{eq_unconst}) as
\begin{equation}
z_{t+1} = G_{\lambda}(z_t) = \begin{cases}G(z_t),~~{\rm with~probability}~\lambda,\\  G_0(z_t),~{\rm with~probability}~1-\lambda,\end{cases}\nonumber
\end{equation}
with $\lambda \in [0, 1]$. The noisy SSR process is the superposition of the SSR process and the unconstrained jump, where the former process occurs with probability $\lambda$.

\begin{figure}[t]
  \centering
  \scalebox{1}{\includegraphics{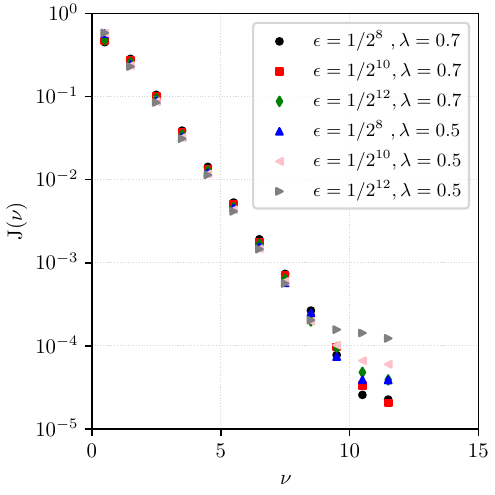}}
  \caption{For the continuous noisy SSR: The data collapse for the universal scaling function for the survival time distribution [cf. Eq.~(\ref{noisy_sur_dis})], with log-linear binning.}
\label{fig4}
\end{figure}

As expected, the mean and variance computed earlier for the discrete noisy SSR~\cite{Yadav_2017_ssr2} should have a similar dependency with the visibility parameter 
\begin{eqnarray}
\langle \tau \rangle \sim \epsilon^{-\theta}~~~~~{\rm and}~~~~
\sigma_{\tau}^2 \sim \epsilon^{-2\theta},
\label{noisy_cumulants}
\end{eqnarray}
with $\theta = 1-\lambda$. And, the survival time distribution satisfies a scaling behavior
\begin{equation}
\mathcal{P}_{\epsilon}(\tau) \sim \epsilon^{\theta}J(\tau\epsilon^{\theta}),
\label{noisy_sur_dis}
\end{equation}
where the universal function, independent of $\lambda$ and $\epsilon$, behaves as $J(\nu) \sim \exp(-\beta\nu)$, where $\beta$ is a constant. Numerical results shown in Figs.~\ref{fig3} and \ref{fig4} are in good agreement with Eqs.~(\ref{noisy_cumulants}) and (\ref{noisy_sur_dis}), which describe the survival time statistics for the continuous noisy SSR process.

\section{Sample space size distribution}{\label{sec_5}}
The multiplicative stochastic processes, even with generalized form, have been found to explain power-law distribution with an additional condition~\cite{Sornette_1997, Takayasu_1997, Susanna_1999, Yamamoto_2012, Yamamoto_2014}. The conditional element could be related to boundary condition, random stopping, or reset. Note that this is one of the crucial mechanisms responsible for the emergence of scale-invariant features in many applications. Both SSR and constrained multiplicative process can explain the fragmentation events.

In the steady state, when the systems' state is trapped in the attractor or simply $z(\tau)\leq\epsilon$, the probability that the noisy SSR process has sample space size $z$ follows a power-law behavior
\begin{equation}
\mathbb{P}_{\epsilon}(z) \sim z^{-\alpha}.
\label{const_size_dist}
\end{equation}
To compute this distribution numerically, we collect samples of $z(\tau)$ for $10^8$ independent realizations of the process. The resolution or visibility parameter $\epsilon$ sets a lower cutoff size.

\begin{figure}[t]
  \centering
  \scalebox{1}{\includegraphics{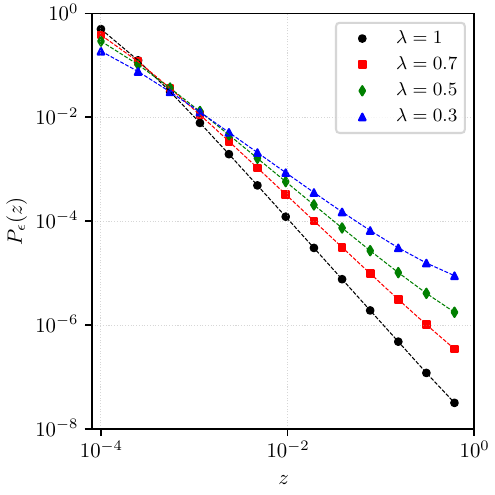}}
  \caption{The plot of the sample size distribution with parameter $\lambda$ [cf. Eq.~(\ref{const_size_dist})]. Here $\epsilon = 10^{-4}$, and the size $z$ is scaled with respect to $\epsilon$, i.e., $z\to z/\epsilon$. The normalized distribution is log binned. The bin width is taken to be $[2^r, 2^{r+1}-1]$, where $r$ runs from 0 to 13, and the size is chosen as the average of lower and upper values of each bin. Again, we scale $z \to \epsilon z$ to compensate for the previous scaling. Using best fit in the range $[10^{-3}, 10^{-1}]$, the estimated exponents for different values of $\lambda = 1, 0.7, 0.5,$ and 0.3 are as follows: $\alpha = 1.98(1)$, $1.68(1)$, $1.48(1)$, and $1.26(1)$, respectively. }
\label{fig5}
\end{figure}

Consider a set of similar objects. An unconstrained stochastic process results when we randomly break each object once. The continuous SSR process emerges as follows: Take a single object and successively break it. Note that the fragmentation depends on visibility parameter. Mixing the two events yields the noisy SSR process, where the SSR process takes place with probability $\lambda$. In the first case, the size distribution is uniform. For the SSR, the distribution behaves as $\mathbb{P}_{\epsilon}(z) \sim z^{-2}$ (cf. Appendix). The SSR process seems to be pulling elements for the dynamical variable towards the visibility parameter. On the other hand, the unconstrained stochastic process introduces a sudden jump between $[\epsilon, 1]$. For the noisy SSR process, the size distribution varies as: $\mathbb{P}_{\epsilon}(z) \sim z^{-\alpha}$. Our simulation results [cf. Fig.~\ref{fig5}] show that the exponent depends upon the parameter $\lambda$ in a simple manner as $\alpha = 1 + \lambda$ [cf. Eq.~(\ref{eq_size_exp})].

We argue to predict the exponent. Let us denote $\mathbb{P}(z,\tau)$, the joint probability density that, after time $\tau$, the size of sample space is $z$. This is expected to be a homogeneous function, since $\langle \tau \rangle \sim z^{-\theta}$, where $\theta = 1 - \lambda$ [cf. Eq.~(\ref{noisy_cumulants}) with $\epsilon \to z$]. Then, introduce the following scaling ansatz \cite{Manchanda_2013}
\begin{equation}
 \mathbb{P}(z,\tau) \sim z^{-(\tau_z-\theta)}H(\tau z^{\theta}).\nonumber
\end{equation}
Clearly, in continuum limit, we get integrating over $\tau$
\begin{equation}
 \mathbb{P}(z) = \int_{0}^{\infty} \mathbb{P}(z,\tau) d\tau \sim z^{-\tau_z}. 
 \label{z_dist_int}
\end{equation}
Using Eqs.~(\ref{z_dist}) and (\ref{z_dist_int}), one identifies $\tau_z = 2$. Thus, $\mathbb{P}(z,\tau) \sim z^{-(1+\lambda)}H(\tau z^{\theta}).$

Since we are eventually interested in computing constrained size distribution, we have
\begin{equation}
\mathbb{P}_{\epsilon}(z) \sim \mathbb{P}(z,\tau = \langle \tau \rangle) \sim z^{-(1+\lambda)}H(\epsilon^{-\theta} z^{\theta}).\nonumber
\end{equation}
Thus, the exponent as a function of the parameter $\lambda$ is given by 
\begin{equation}
\alpha = 1 + \lambda,~~~{\rm with}~~~\lambda \in [0, 1].
\label{eq_size_exp}
\end{equation}

Ref.~\cite{Blasius_2009} provides an explicit illustration for a related problem describing popularity in chess openings. For $\lambda = 1$, the exponent $\tau_z = 2$ is  exact, and also robust even if $p(x)$ follows a nonuniform distribution of a power-law type: $p(x) = (1+\gamma)x^{\gamma}$, which is uniform for $\gamma = 0$.

\section{Summary}{\label{sec_6}}
To summarize, we have presented a simple model for the continuous version of the SSR process, inspired by the description of fragmentation events. Moreover, the model also represents a constrained multiplicative stochastic process, one of the key mechanisms explaining the origin of scale-invariant features observed in complex systems. One can easily generalize the model to the continuous noisy SSR process, explaining the variability of the exponent associated with Zipf's law.

Qualitatively, the survival time statistics for the process behave similarly to that of the discrete version. The difference between mean and variance of the survival time differs from the discrete SSR process. One of the striking observations is that the constrained size of sample space exhibits a power-law distributed probability, with an exponent in the range 1 to 2. The exponent range is physically relevant, particularly in the context of fragmentation processes (cf. Refs.~\cite{Morgado_2007, Ramola_2008, Hermann_2010, Astrom_2006, Fytas_2015}). It would be further interesting to examine the extent of scaling behavior in the size of sample space in the generalized SSR.

\section*{ACKNOWLEDGMENT}
RC acknowledges UGC, India, for financial support through a Senior Research Fellowship.

\appendix*{\label{app_1}}
\section{The probability density for the size of sample space without constraint}
The evolution of the size of sample space follows Eq.~(\ref{zt_update}), where $x_i$ is a uniformly distributed random variable in the interval [0, 1]. Here, our primary goal is to determine $\mathbb{P}(z)$, the probability density for the size of sample space. We can compute this by using 
\begin{equation}
\mathbb{P}(z) = \sum_{\tau}\mathbb{P}_{\tau}(z),
\label{size_dist}
\end{equation}
where $\mathbb{P}_{\tau}(z)$ is the probability density of the size when the process is stopped after time $\tau$. The sum is carried over all possible values of $\tau $ running from 0 to $\infty$. Note that the size conservation \cite{Blasius_2009} suggests that 
\begin{equation}
\int_{0}^{1} z \mathbb{P}_{\tau}(z) dz= 1.
\label{eq_norm}
\end{equation}
Eq.~(\ref{eq_norm}) implies that $\mathbb{P}_{\tau}(z)$ is not normalized. This can be fixed by introducing a normalized size density function 
\begin{equation}
f_{\tau}(z) = z \mathbb{P}_{\tau}(z).
\label{norm_z_dist}
\end{equation}

Since it is preferable to work with sum of random variables rather than multiplication, introducing $y = \ln 1/z$ and $\xi_i = -\ln x_i$, Eq.~(\ref{zt_update}) reduces
\begin{equation}
y = \sum_{i=0}^{\tau}\xi_i. \nonumber
\end{equation}
It can be easily checked that $\xi$ satisfies an exponential density function $p(\xi) = \exp(-\xi)$. In turn, the sum of $\tau$ independent identically and exponentially distributed random variables follows a gamma distribution \cite{Feller_1971} 
\begin{equation}
f_{\tau}(y) = \frac{1}{(\tau-1)!}y^{\tau-1}\exp(-y).
\label{gamma_eq}
\end{equation}

For the transformation $z = \exp(-y)$, Eq.~(\ref{gamma_eq})  immediately leads to normalized log-gamma density function 
\begin{equation}
f_{\tau}(z) = \frac{1}{(\tau -1)!}\left(\ln \frac{1}{z}\right)^{\tau-1}.
\label{log_gamma}
\end{equation}
Also, note that $f_{\tau}(z)$ satisfies a recursion relation as
\begin{equation}
f_{\tau}(z) = \int_{z}^{1}f_{\tau-1}\left(\frac{z}{x}\right)\frac{dx}{x}.\nonumber
\end{equation}
Since $\sum_{\tau}f_{\tau}(z) = 1/z$, using Eqs.~(\ref{size_dist}), (\ref{norm_z_dist}), and (\ref{log_gamma}), it is easily seen that 
\begin{equation}
 \mathbb{P}(z) = \frac{1}{z^2}.
 \label{z_dist}
\end{equation}

\bibliography{ver4}
\bibliographystyle{myrev}

\end{document}